\documentclass[aps,pra,twocolumn,groupedaddress]{revtex4}

\usepackage{graphicx}
\usepackage{bm}
\begin{document}

%\preprint{}

\title{Magnetic trapping of metastable $^3P_2$ atomic
strontium}


\author{S. B. Nagel, C. E. Simien, S. Laha, P. Gupta, V. S. Ashoka, and T. C. Killian}
%\thanks{}
\affiliation{Rice University, Department of Physics and Astronomy
and Rice Quantum Institute, Houston, Texas, 77251}


\date{\today}

\begin{abstract}
We report the magnetic trapping of metastable $^3P_2$ atomic
strontium. Atoms are cooled in a magneto-optical trap (MOT)
operating on the dipole allowed $^1S_0-^1P_1$ transition at 461
nm. Decay via $^1P_1\rightarrow {^1D_2}\rightarrow {^3P_2}$
continuously loads a magnetic trap formed by the quadrupole
magnetic field of the MOT. Over $10^8$ atoms at a density of $8
\times 10^9$~cm$^{-3}$ and temperature of 1~mK are trapped.
%The number of atoms is limited by the loading rate of
%$10^9$~atoms/s and background gas collision rate of $2$~$s^{-1}$.
The atom temperature is significantly lower than what would be
expected from the kinetic and potential energy of atoms as they
are transferred from the MOT. This suggests that thermalization
and evaporative cooling are occurring in the magnetic trap.
\end{abstract}

\pacs{32.80.Pj}
%\keywords{}

\maketitle

%\section{Introduction\label{introduction}}
%\subsection{}
%\subsubsection{}

% If in two-column mode, this environment will change to single-column
% format so that long equations can be displayed. Use
%\begin{widetext}
%\end{widetext}

Laser-cooled alkaline-earth atoms
%, magnesium, calcium, and strontium,
offer many possibilities for practical applications and
fundamental studies. The two valence electrons in these systems
give rise to triplet and singlet levels connected by narrow
intercombination lines that are utilized for optical frequency
standards \cite{obf99}. Laser cooling on such a transition in
strontium may lead to a fast and efficient route to all-optical
quantum degeneracy \cite{kii99,iik00}, and there are abundant
bosonic and fermionic isotopes to use in this pursuit. The lack of
hyperfine structure in the bosonic isotopes and the closed
electronic shell in the ground states make alkaline-earth atoms
appealing testing grounds for cold-collision theories
\cite{mjs01,zbr00,dva99}, and collisions between metastable
alkaline-earth atoms is a relatively new and unexplored area for
research \cite{der01}.
%that can display interesting physics, as
%demonstrated with Bose-Einstein condensates of metastable helium
%\cite{rsb01,slw01}.% in a similar metastable state.

In this paper we characterize a technique that should benefit all
these experiments - the continuous loading of metastable $^3P_2$
atomic strontium ($^{88}$Sr) from a magneto-optical trap (MOT)
into a purely magnetic trap.  This idea was discussed in a recent
theoretical study of alkaline-earth atoms and ytterbium
\cite{lbm02}. Katori {\it et al}. \cite{kii01} and Loftus {\it et
al}. \cite{lxh02} have also reported observing this phenomenon in
their strontium laser-cooling experiments. Continuous loading of a
magnetic trap from a MOT was recently described for chromium atoms
\cite{ssh01}.

%This paper represents the first experimental study of this process
%for alkaline-earth atoms.
This scheme should allow for collection  of large numbers of atoms
at high density since atoms are shelved in a dark state and less
susceptible to light-assisted collisional loss mechanisms
\cite{kdj93,mjs01,dva99}. It is an ideal starting place for many
experiments such as sub-Doppler laser-cooling on a transition from
the metastable state, as has been done with calcium \cite{ghe02},
production of ultracold Rydberg gases \cite{rtn00} or plasmas
\cite{kkb99}, and evaporative cooling to quantum degeneracy.
%Like Bose-Einstein condensates of helium \cite {rsb01,slw01} in a
%similar metastable state, the
%$^3P_2$ Sr atoms may possess cold collision properties that
%enhance the stability of a metastable condensate \cite{der01}.
Optical frequency standards based on laser-cooled alkaline-earth
atoms, which are currently limited by high sample temperatures
\cite{obf99},
 %and the resulting residual Doppler shifts,
may benefit from the ability to trap larger numbers of atoms and
evaporatively cool them in a magnetic trap.

\begin{figure}
%print to pdfwriter from ppoint,open resulting pdf with ghostview, convert to
%eps with pswrite, max resoltuion
  %\includegraphics[width=3in,clip=true,trim=100 290 50 80 ]{energydiagramstraight.eps}\\
  \includegraphics[width=3in,clip=true,trim=100 290 50 80 ]{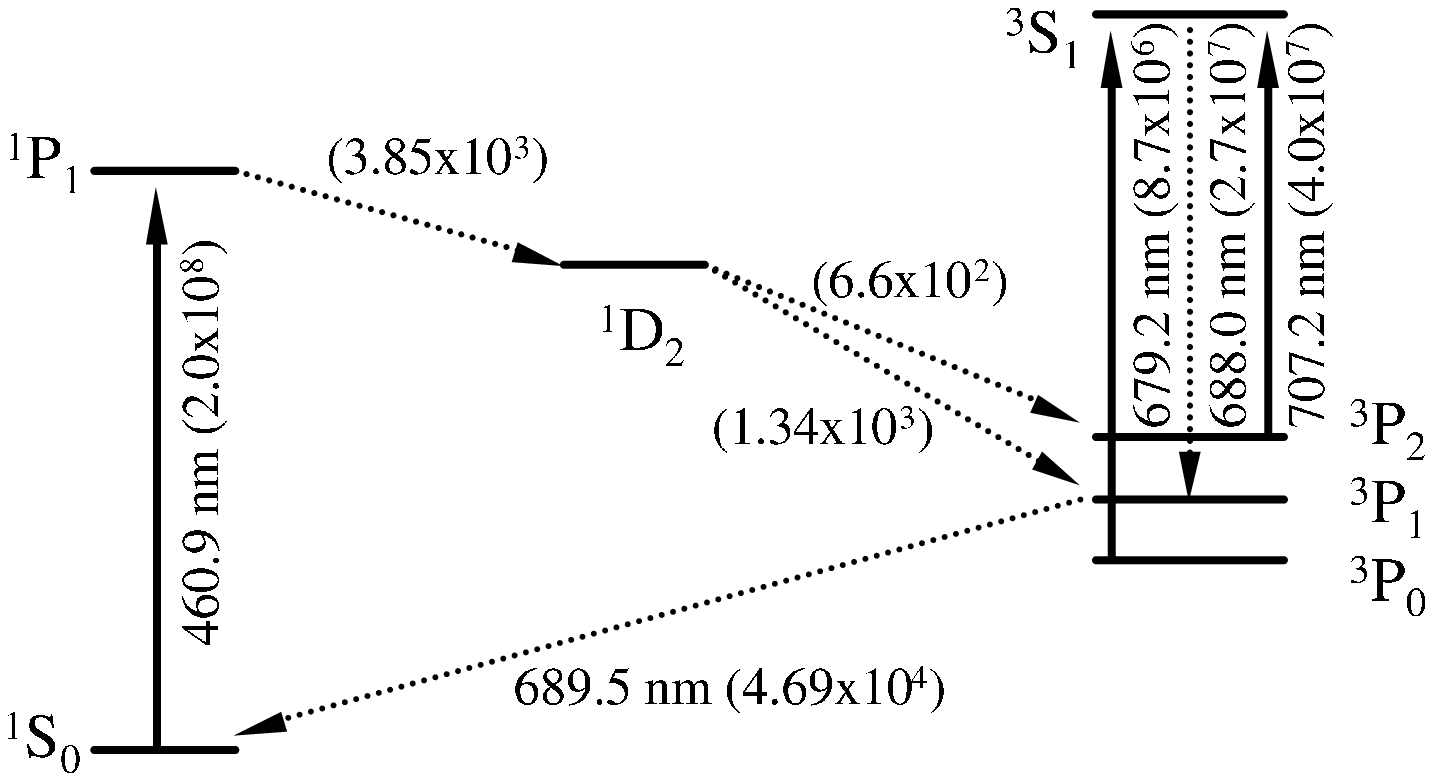}\\
  \caption{Strontium energy levels involved in the trapping of $^3P_2$ atoms.
  Decay rates (s$^{-1}$)
   and selected excitation wavelengths are given.
  Laser light used for the experiment is indicated by solid lines.}\label{energydiagram}
\end{figure}

%\section{Trapping $^3P_2$ strontium\label{Trapping}}
We will first describe the operation of the Sr MOT and how this
loads the magnetic trap with $^3P_2$ atoms. Then we will
characterize the loading and decay rates of atoms in the magnetic
trap. Finally, we will present measurements of the $^3P_2$ sample
temperature.

%For the experiment described here,
Sr atoms are loaded from a
Zeeman-slowed atomic beam \cite{pmp85} and cooled and trapped in a
standard MOT \cite{rpc87}. Both operate on  the 461 nm {$^1S_0-
{^1P_1}$} transition (Fig.\ \ref{energydiagram}). Blue light is
generated by frequency doubling the output of a Ti:sapphire laser
using KNbO$_3$ in two external buildup cavities \cite{bft97}. 150
mW of power, red-detuned from resonance by 585~MHz, is available
for the Zeeman slower.
%and the MOT light
%is 57~MHz detuned below atomic resonance.
Three beams of about 1~cm diameter, each with intensity $I\le
I_{sat}=45$~mW/cm$^2$, and 57~MHz red-detuned, are sent to the
apparatus and retroreflected to produce the 6-beam MOT.

The 30~cm long Zeeman slower connects a vacuum chamber for the Sr
oven and nozzle to the MOT chamber. Each chamber is evacuated by a
75~l/s ion pump. When the Sr oven is operated at its normal
temperature of about $550 ^{\circ}$C, the pressure in the MOT
chamber is about $5 \times 10^{-9}$~torr, and the oven chamber is
at $4 \times 10^{-8}$~torr.

Extended cavity diode lasers
%serve as repump lasers for the
at 679 and 707 nm remove population from the $^3P_0$ and $^3P_2$
levels. Each laser provides several hundred $\mu$W of power and is
locked to an absorption feature in a discharge cell. These are not
used for operation of the MOT during experiments with $^3P_2$
atoms. They serve to repump atoms from the $^3P_2$ level to the
ground state via the $^3S_1$ and $^3P_1$ levels for imaging
diagnostics.

The quadrupole magnetic field for the MOT is produced by flowing
up to 80~A of current  in opposite directions through two coils of
36 turns each, with coil diameter of 4.3~cm and separation of
7.7~cm. The maximum current produces a field gradient along the
symmetry axis of the coils of 115~G/cm. Such a large field
gradient, about $10\times$ the norm for an alkali atom MOT, is
required because of the large decay rate of the excited state
($\Gamma_{^1P_1}=2\pi \times 32$~MHz) and the comparatively large
recoil momentum of 461 nm photons.

Typically $10^7-10^8$ atoms are held in the MOT, at a peak density
of $n \approx 1\times 10^{10}$~cm$^{-3}$,  with an rms radius of
1.2~mm and temperatures from $2-10$~mK. The cooling limit for the
MOT is the Doppler limit ($T_{Doppler}=0.77$~mK) because the
ground state lacks degeneracy and thus cannot support sub-Doppler
cooling. Higher MOT laser power produces higher MOT temperature,
but also a larger number of trapped atoms. These sample parameters
are measured with absorption imaging of a near resonant probe
beam. The temperature is determined by monitoring the velocity of
ballistic expansion of the atom cloud \cite{wrs89} after the trap
is extinguished.

Atoms escape from the MOT due to {$^1P_1 -{^1D_2}$} decay
%($\Gamma_{^1P_1-^1D_2}=2\pi \times 600$~Hz)
as discussed in \cite{lbm02}. From the {$^1D_2$} state atoms
either decay to the $^3P_1$ state and then to the ground state and
are recaptured in the MOT, or they decay to the $^3P_2$ state,
which has a lifetime of 17 min \cite{der01}. The decay rates are
given in Fig.\ \ref{energydiagram}. The resulting MOT lifetime of
$11-55$~ms was measured by turning off the Zeeman slowing laser
beam and monitoring the decay of the MOT fluorescence. The
lifetime is inversely proportional to the fraction of time atoms
spend in the $^1P_1$ level, which varies with MOT laser power.
Light-assisted collisional losses from the MOT \cite{dva99} are
negligible compared to the rapid {$^1P_1 -{^1D_2}$} decay.
%If
%the repump lasers are applied to the MOT, the lifetime increases
%by over an order of magnitude and is limited by light assisted two
%body collisions \cite{dva99}.

 The  {$m_j=2$} and {$m_j=1$}~{$^3P_2$} states can be trapped in
the MOT quadrupole magnetic field. Such a quadrupole magnetic trap
was used for the first demonstration of magnetic trapping of
neutral atoms \cite{mpp85}, but in that case atoms were loaded
directly from a Zeeman-slowed atomic beam.

Near the center of the trap, the magnetic interaction energy for
$^3P_2$ atoms takes the form
\begin{equation}\label{interactionenergy}
    U_{m_j}=-{\bm{\mu}}_{m_j} \cdot  \textbf{B}=g \mu_B m_j\hspace{.015in} b
\sqrt{x^2/4 + y^2 + z^2/4},
\end{equation}
where $m_j$ is the angular momentum projection along the local
field, $g=3/2$ is the g-factor for the $^3P_2$ state, $\mu_B$ is
the Bohr magneton, and $b \le 115$~G/cm is the gradient of the
magnetic field along the symmetry (y) axis of the quadrupole coil.
For the $m_j=2$ state and  the maximum $b$,  $g \mu_B m_j /k_B=
200$~$\mu$K/G and the barrier height for  escape from the center
of the magnetic trap is $15$~mK. Gravity, which is oriented along
$z$, corresponds to an effective field gradient of only 5 G/cm for
Sr and is neglected in our analysis.

\begin{figure}
  \includegraphics[width=3.3in,clip=true]{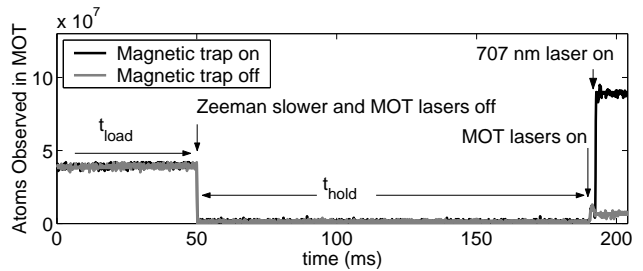}\\
  \caption{Magnetic trapping of $^3P_2$ atoms. Ground state
  atoms are detected by fluorescence from the MOT lasers.
  If  the quadrupole magnetic field is left on during
  $t_{hold}$ (black trace), large numbers of $^3P_2$ atoms are magnetically trapped until
  the 707
  nm laser returns them to the ground state.
  The residual fluorescence after $t_{hold}$ for the gray trace arises from
  background scatter off atoms in the atomic beam. Reloading of the MOT from the atomic beam
  is negligible with the Zeeman slower light blocked. There is a 1 ms delay
  between the
  MOT and 707 nm laser turn on to allow the
  %servo controls for the
  MOT light
intensity to reach a stable value.}\label{timeline}
\end{figure}

Typical data showing the magnetic trapping is shown in Fig.\
\ref{timeline}. We are unable to directly image atoms in the
$^3P_2$ state, so we use the 679 and 707 nm lasers to repump them
to the ground state for fluorescence detection on the
$^1S_0-^1P_1$ transition. The details are as follows. The MOT is
operated for $t_{load}\le 1300$~ms during which time atoms
continuously load the magnetic trap.
%Fluorescence provides a measure of the number of atoms in the MOT.
The MOT and Zeeman slower light is then extinguished, and after a
time $t_{hold}$, the MOT lasers and repump laser at  707~nm are
turned on.  The 679 nm laser is left on the entire time. Any atoms
in the $^3P_2$ state are cycled through the $^3P_1$ level to the
ground state within $500$~$\mu$s of repumping, and they fluoresce
in the field of the MOT lasers. If the magnetic field is not left
on during $t_{hold}$, Fig.\ \ref{timeline} shows that a negligible
number of ground state atoms are present in the MOT when the
lasers are turned on. If the magnets are left on, however, the MOT
fluoresence shows that {$^3P_2$}
 atoms were held in the magnetic
trap.

\begin{figure}
  %\includegraphics[width=2.5in]{metadecayVpressure.eps}
  %  \includegraphics[width=2.5in,trim=-0 0 5 0]{metaloadVmotlossnofit.eps}
    \includegraphics[width=2.5in]{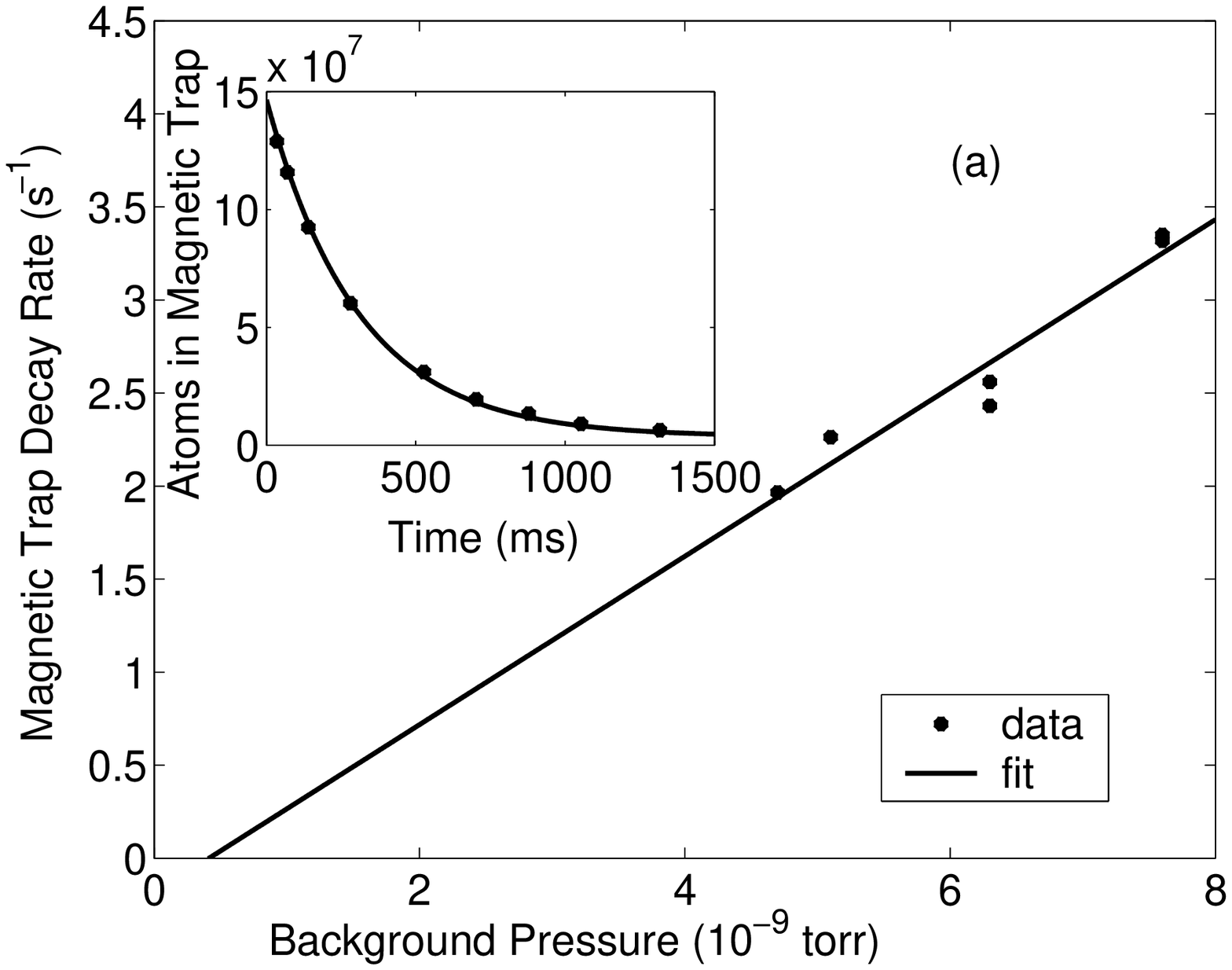}
    \includegraphics[width=2.5in,trim=-0 0 5 0]{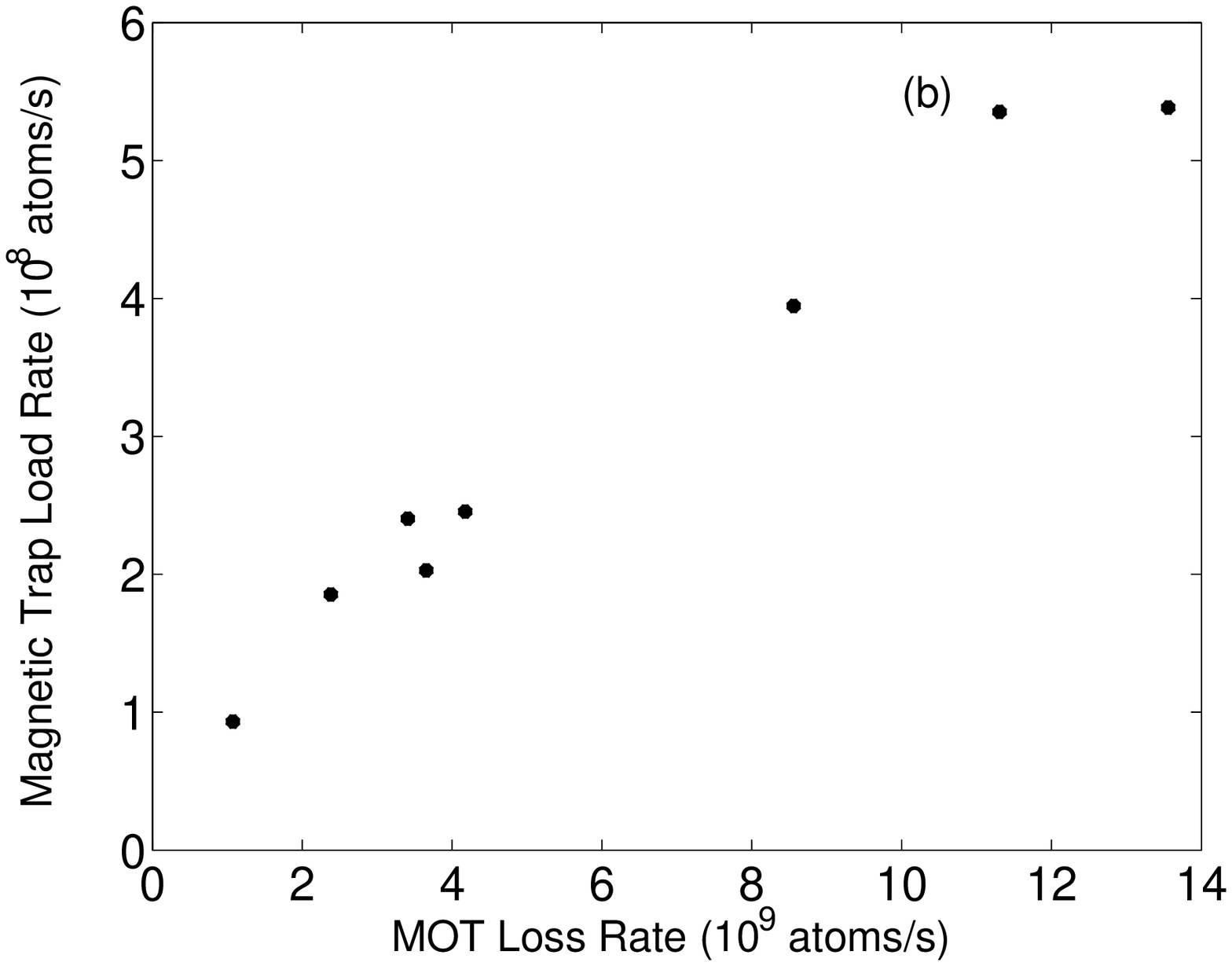}
  \caption{(a) The lifetime of atoms in the magnetic trap is limited
  by collisions with background gas molecules. The linear fit
  extrapolates to zero at zero pressure within statistical
  uncertainties.
  Inset: A typical fit of the decay
  of the number of trapped atoms
  to a single exponential.
  (b) The magnetic trap loading rate is plotted against the
    MOT loss rate. Data correspond to various MOT laser powers
    and  slow-atom fluxes from the atomic beam.
  %The linear fit is to loading rates below $3 \times 10^{8}$~atoms/s
  }
  \label{metadecayVpressure}
  \label{metaloadVmotloss}
\end{figure}

The maximum number of $^3P_2$ atoms trapped is about $1 \times
10^8$, and the peak density is about $8 \times 10^{9}$~cm$^{-3}$.
To determine what limits this number, we varied $t_{hold}$ and saw
that the number of $^3P_2$ atoms varied as $N_0\rm{e}^{-\gamma
t_{hold}}$. The fits were excellent and the decay rate was
proportional to background pressure as shown in Fig.\
\ref{metadecayVpressure}a. This implies that for our conditions,
the trap lifetime is limited by collisions with residual
background gas molecules, and strontium-strontium collisional
losses are not a dominant effect.
%The decay rate was proportional to background pressure as shown in
%Fig.\ \ref{metadecayVpressure}. The fits were excellent for times
%longer than 10~ms, which, along with the good correlation with
%background pressure, implies that for densities below $5 \times
%10^{-9}$, the decay
% is dominated by collisions with the background gas.
%The data shows some large loss rate at $t_{hold}<10$~ms, but no
%correlation was found with atom density. This could be the time
%required for MOT atoms to leave the large MOT recapture volume,
%but more study is required to preclude a many-body collisional
%loss process in the magnetic trap at short times.

The magnetic trap loading rate was determined by holding
$t_{hold}$ constant and varying $t_{load}$.
%The data fit a single exponential loading curve well.
The loading rate correlates with the  atom loss rate from the MOT
(Fig.\ \ref{metaloadVmotloss}b). At low MOT loss rates about 10\%
of the atoms lost from the MOT are captured in the magnetic trap.
From the Clebsch-Gordon coefficients involved in atom decay from
the $^1P_1$ state, and the magnetic sublevel distribution for
atoms in the MOT, one expects that about 25\% of the atoms
decaying to $^3P_2$ enter the $m_j=1$ or $m_j=2$ states. This is
significantly higher than the largest observed efficiencies, and
we may be seeing signs of other processes, such as losses due to
collisions with MOT atoms. This phenomenon dominated dynamics
during loading of a magnetic trap from a chromium MOT
\cite{ssh01}.

At larger MOT loss rates (corresponding to higher MOT laser
intensities,  MOT temperatures, and trapped $^3P_2$ atom
densities), the efficiency of loading the magnetic trap decreases
by about a factor of two. %$^3P_2$
MOT temperatures approach the trap depth for the largest loading
rates and we attribute the decreasing efficiency to escape of
atoms over the magnetic barrier.

%\section{$^3P_2$ Temperature\label{$^3P_2$ Temperature}}

\begin{figure}
  %\includegraphics[width=3.3in,clip=true]{metapicxy.eps}\\
  \includegraphics[width=3.3in,clip=true]{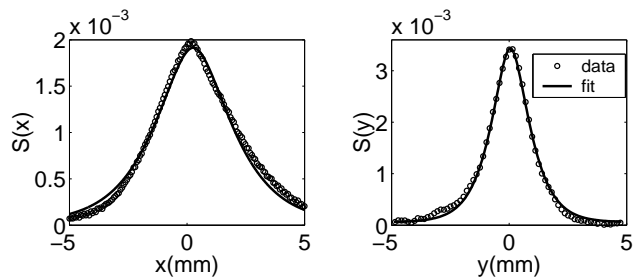}\\
  \caption{Distributions of $^3P_2$ atoms extracted from absorption
  images of ground state atoms shortly after repumping. The fits, which assume
  thermal equilibrium and a pure sample of $m_j=2$ atoms,  yield
  number ($1.2 \times 10^8$), peak density ($8 \times 10^9$~cm$^{-3}$),
  and temperature (1.3~mK) of the atoms.}\label{combinedpic}
\end{figure}

The 500~$\mu$s required for repumping is fast compared to the
timescale for motion of the atoms, so absorption images of ground
state atoms immediately after repumping provides a measure of the
density distribution of the magnetically trapped sample. For these
measurements, the magnetic trap is loaded for $1.3$~s. Then the
magnetic field is turned off and the repump lasers are turned on.
After
%The MOT lasers are never turned on.
500~$\mu$s,  an 80~$\mu$s pulse of a weak 461 nm probe beam ($I\ll
I_{sat}$), 12.5~MHz detuned below resonance, illuminates the atom
cloud and falls on a CCD camera. We record an intensity pattern
with atoms present, $I(x,y)_{atoms}$, and a background pattern
with no atoms present, $I(x,y)_{back}$.
%\begin{equation}\label{Beerslaw}
%    I(x,y)_{atoms}=I(x,y)_{back}{\rm exp}[-\sigma_{abs}
%    \int_{-\infty}^{\infty} dz \hspace{.025in} n(x,y,z)],
%\end{equation}
%
To analyze the data, we plot
\begin{eqnarray}\label{SX}
    S(x)&=&\int_{image}dy \hspace{.025in} {\rm ln} [I(x,y)_{back}/I(x,y)_{atoms}] \nonumber \\
    &=&\sigma_{abs}\int_{image}dy \int_{-\infty}^{\infty} dz \hspace{.025in}
    n(x,y,z),
\end{eqnarray}
and the analogously defined S(y), where $\sigma_{abs}$ is the
absorption cross section % removed at the request of reviewer "at this detuning"%
and $n(x,y,z)$ is the atom density (Fig.\ \ref{combinedpic}).
Because we do not know the distribution of magnetic sublevels, we
make the simplifying assumption that all atoms are in the $m_j=2$
state, and the density is given by
\begin{equation}\label{Beerslaw2}
    n(x,y,z)=n_{0}{\rm exp} [-U_{2}(x,y,z)/k_B T],
\end{equation}
A numerical approximation to Eq.\ \ref{SX} fits the data very
well.

%\begin{equation}\label{Beerslaw2}
%    I(x,y)_{atoms}=I(x,y)_{back}\rm{exp}\{-\sigma_{abs}
%    \int_{-\infty}^{\infty} dz \Sigma_{m_j=1,2} n_{m_j}\rm{exp} [-U_{m_j}(x,y,z)/k_B
%    T_{m_j}]\},
%\end{equation}
%where  $\sigma_{abs}$ is the absorption cross section at this
%detuning, $n_{m_j}$, $T_{m_j}$, and $U_{m_j}$ are the peak
%intensity, temperature, and  potential (Eq.\
%\ref{interactionenergy}) for the $m_j$ state. We are assuming
%thermized populations.

%The distribution of magnetic sublevels can be determined with a
%Stern-Gerlach type experiment, and will be the subject of a future
%study. But
Our assumption for magnetic sublevel distribution means that the
extracted temperatures are upper bounds, but one would expect the
$m_j=1$ level to be less populated. Due to the smaller magnetic
moment, the trapping efficiency for $m_j=1$ atoms decreases
substantially as the MOT temperature increases, dropping by about
a factor of 5 for a MOT temperature of 12~mK compared to only a
factor of two for $m_j=2$ atoms. Atoms with $m_j=1$ can also be
lost from the trap through spin-exchange collisions, which are
typically rapid in ultracold gases. Calculated rates for spin
exchange collisions for alkali atoms in magnetic traps are
typically $10^{-11}$~cm$^3$/s, although they can approach
$10^{-10}$~cm$^3$/s \cite{kbv97}.
%removed at the request of the reviewer " which often leads to spontaneous
%spin-polarization in dense, ultracold atomic samples."%

%The assumption of thermal equilibrium is supported by the quality
%of the fits.
%%Peak densities extracted from the fits approach
%%$10^{10}$~cm$^{-3}$.
%For $1$~mK atoms at a peak density of $10^{10}$~cm$^{-3}$,
%thermalization on a one second timescale requires an average
%elastic cross larger than $10^{-11}$~cm$^{2}$. This is large, but
%not unreasonable for a cold collision. It is not possible to
%extract meaningful estimates of scattering lengths because of the
%uncertainty in the magnetic sublevel distribution in the trap and
%the possibility that several partial waves are contributing to
%collisions.

We have assumed thermal equilibrium in our analysis, but this is
reasonable.
%Peak densities extracted from the fits approach
%$10^{10}$~cm$^{-3}$.
Thermal equilibration would need to occur on less than a few
hundred ms timescale. Using a recently calculated s-wave elastic
scattering length for $^3P_2$ atoms of $a=6$~nm \cite{dpk02}, the
collision rate for identical atoms is $n v 8 \pi a^2\approx
9$~s$^{-1}$ for $n=10^{10}$~cm$^{-3}$ and $v=\sqrt{2 k_B
T/M}=1$~m/s ($T=3$~mK).

%\!\(N[\(10\^16\) 1\ 8\ \[Pi]\ \((115\ 5.3\ 10\^\(-11\))\)\^2]\)
%9.33
%\!\(N[115\ 5.3\ 10\^\(-11\)]\)
%\!\(6.095`*^-9\)

\begin{figure}
  %\includegraphics[width=3in]{metatempinset.eps}\\
  \includegraphics[width=3.3in]{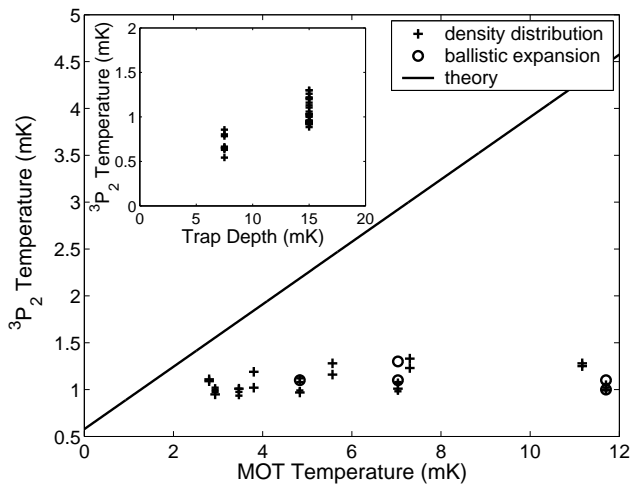}\\
  \caption{The $^3P_2$ temperature
  %,extracted from fits such as
  %in Fig.\ \ref{metapic550c2.0v80aax} and from ballistic expansion of atoms after release
  %from the trap,
  is significantly lower than expected from a
  simple model that is described in the text.
  The inset shows that
  the temperature tracks the magnetic trap depth, as expected for
  evaporative cooling.
  The scatter of the temperature measurements is characteristic of our
  statistical
  uncertainty, and there is a scale uncertainty of $~25$\% due to
  calibration of the imaging system.  The magnetic trap depth is 15~mK
  for the main figure.
  %(Other hypothesis - more likely to trap atoms that decay near
  %trap center.)
   }\label{motmetatemps.eps}
\end{figure}

The most interesting parameter obtained from the fits is the
temperature, which is plotted in Fig.\ \ref{motmetatemps.eps} as a
function of MOT atom temperature. The values are significantly
colder than what one would expect from a simple theory developed
in \cite{ssh01} and plotted in the figure. The expected
temperature is determined by assuming the kinetic energy and
density distribution in the MOT are preserved as atoms decay to
the metastable state. The $^3P_2$ potential energy distribution is
then given by the magnetic trap potential energy corresponding to
the density distribution of the MOT.
% The values are significantly
%colder than what one would expect by assuming the temperature is
%determined by the kinetic energy in the MOT and the potential
%energy in the magnetic trap corresponding to the density
%distribution of the MOT \cite{ssh01}, which is the theory line in
%the figure.
Actual $^3P_2$ atom temperatures cluster around 1~mK for a 15~mK
trap depth, while the expected temperature approaches
 4.5~mK for
the hottest MOT conditions. We confirmed these measurements  by
determining the $^3P_2$ atom temperature from ballistic expansion
velocities, as is done to measure the MOT temperature.

As shown in the inset of Fig.\ \ref{motmetatemps.eps}, the
temperature decreases with decreasing trap depth as would be
expected for evaporative cooling of the sample \cite{kvd96}. For
this data, the magnetic trap depth is held constant during the
entire load and hold time. % cut and reworded as suggested "To confirm this explanation for the low
%atom temperatures, it would be ideal to measure"$
Confirmation of this explanation could be achieved with
measurement of the collision cross section and thermalization rate
in the trap.
%, and determine
%what heating mechanisms balance evaporation to produce
%temperatures of between 1/10 and 1/15 of the trap depth.
We plan
on pursuing these experiments.

If evaporative cooling is working efficiently, it should be
possible to use radio-frequency-induced forced evaporative cooling
to further cool the sample and increase the density. Majorana
spin-flips \cite{mpp85} from trapped to untrapped magnetic
sublevels at the zero of the quadrupole magnetic field will
eventually limit the sample lifetime,
%at at the field yield a lifetime of 1000~s at 1 mK.
but it will still be 10 s at 100 uK. For studies of quantum
degeneracy, the sample would have to be transferred to a magnetic
trap without a field zero or to an optical dipole trap.
Straightforward improvement of our vacuum should yield atom
numbers and densities an order of magnitude higher than currently
attained, and allow us to fully explore potential gains through
evaporative cooling.

This research was supported by the Office for Naval Research,
Research Corporation, Sloan Foundation, and Department of Energy
Office of Fusion Energy Sciences.

\bibliography{bibliography}

\end{document}